\newcommand{\commentt}[2]{#1}
\newcommand{\comment}[1]{}
\newcommand{\swap}[2]{\commentt{#2#1}{#1#2}}

\newcommand{\var}{{\rm Var}}
\newcommand{\std}{{\rm Std}}
\commentt{
\newcommand{\citet}{\citeasnoun}
\documentclass[a4paper,11pt]{article}
\usepackage{graphicx,slashbox,amsthm}
\usepackage[full]{harvard}

\usepackage[ps,dvips]{xy}
\title{General multilevel Monte Carlo methods for pricing discretely monitored
Asian options}
\author{Nabil Kahal\'e
\thanks{\emph{ Europe,  Labex ReFi and Big data research center, 75011 Paris,
France; {e-mail: }{nkahale@escpeurope.eu}.}}
}
\date{\today}
\usepackage{amstext}
\usepackage{amssymb}
\usepackage{amsmath}
\usepackage{setspace}
\usepackage[margin=1in]{geometry}
\usepackage[english]{babel}
\bibliographystyle{dcu}
\usepackage{color}
\usepackage{textcomp}
\usepackage{varioref}
\usepackage{hyperref}

\begin{document}

\newtheorem{example}{Example}[section]
\newtheorem{theorem}{Theorem}[section]
\newtheorem{conjecture}{Conjecture}[section]
\newtheorem{lemma}{Lemma}[section]
\newtheorem{proposition}{Proposition}[section]
\newtheorem{remark}{Remark}[section]
\newtheorem{corollary}{Corollary}[section]
\newtheorem{definition}{Definition}[section]
\numberwithin{equation}{section}
\maketitle
\newcommand{\ABSTRACT}[1]{\begin{abstract}#1\end{abstract}}
\newcommand{\citep}{\cite}
}
{
\documentclass[opre,nonblindrev]{informs3} 
\usepackage[margin=1in]{geometry}
\usepackage{textcomp}
\usepackage{pgfplots}
\DoubleSpacedXII



\usepackage[round]{natbib}
 \bibpunct[, ]{(}{)}{,}{a}{}{,}%
 \def\bibfont{\small}%
 \def\bibsep{\smallskipamount}%
 \def\bibhang{24pt}%
 \def\newblock{\ }%
 \def\BIBand{and}%

\TheoremsNumberedThrough     
\ECRepeatTheorems

\EquationsNumberedThrough    

\MANUSCRIPTNO{} 

\renewcommand{\qed}{\halmos}

\begin{document}
\RUNAUTHOR{Kahal\'e}
\RUNTITLE{Multilevel  methods for discrete Asian options}
\TITLE{General multilevel Monte Carlo methods for pricing discretely monitored
Asian options}
\ARTICLEAUTHORS{%
\AUTHOR{Nabil Kahal\'e}

\AFF{ESCP Europe, Labex Refi and Big data research center, 75011 Paris, France, \EMAIL{nkahale@escpeurope.eu} \URL{}}
} 
\bibliographystyle{informs2014} 
}

\ABSTRACT{
We  describe  general multilevel Monte Carlo methods that estimate the price of an Asian option  monitored at \(m\) fixed dates.  Our approach   yields
  unbiased estimators  with standard deviation \(O(\epsilon)\) in \(O(m+\epsilon^{-2})\) expected time for a variety of processes including the  Black-Scholes model,      Merton's jump-diffusion model, the Square-Root diffusion model, Kou's
double exponential jump-diffusion model, the variance gamma and
 NIG   exponential Levy processes and, via the Milstein scheme,  processes driven by scalar stochastic differential equations. Using the Euler scheme,  our approach  estimates the Asian option price with root mean square error \(O(\epsilon)\)   in   \(O(m+(\ln(\epsilon))^{2}\epsilon^{-2})\) expected time for processes driven by multidimensional stochastic differential equations.
 Numerical experiments  confirm that our approach outperforms the conventional Monte Carlo method by a factor of order \(m\).
     }
\commentt{
}{\maketitle
Subject classifications: Finance: asset pricing; Simulation: Efficiency ; Analysis of algorithms: Computational complexity}
\section{Introduction}
Asian options are financial derivatives whose payoff  depends on the
arithmetic average of an underlying  during a specific time-period.
Asian options are useful to corporations which are exposed to average
exchange rates   or commodity prices over a certain period of time.  Pricing Asian options has been the subject of many studies. Under      the Black-Scholes model, the price of a continuously sampled  Asian option can be expressed as an infinite series~\citep{LinetskyAsian2004}. 
Transform based methods have been used to value Asian options  under Markov
processes~\citep{Cai2015,cui2018single}. A convex programming method that
computes optimal model-independent     bounds on Asian option prices is described in~\citep{kahale2017}. Monte Carlo methods can  price Asian options under various models, but conventional Monte Carlo algorithms have a high computational cost, which motivates the need to improve the efficiency of such methods. Control variate techniques for pricing Asian options with Monte Carlo simulation are given in~\citep{kemna1990pricing,dingecc2012,shiraya2017general}. An importance sampling algorithm for pricing Asian options is derived in~\citep{glasserman1999asymptotically}. When the underlying follows a stochastic differential equation (SDE) satisfying certain regularity conditions, the multilevel Monte Carlo method (MLMC)  described in~\citep{Giles2008} estimates the price  of a continuously monitored Asian option with mean square error \(\epsilon^{2}\)  in \(O((\ln(\epsilon))^{2}\epsilon^{-2})\) time using the Euler  discretization. This computational cost has been reduced to \(O(\epsilon^{-2})\) time     using the Milstein scheme for scalar SDEs~\citep{giles2008improved,giles2013numerical} and multi-dimensional  SDEs~\citep{giles2014antithetic}. For a broad class of pure-jump exponential Levy processes, the MLMC method described in~\citep{giles2017Levy} estimates the price of  a continuously monitored Asian option with mean square error \(\epsilon^{2}\)   in   \(O(\epsilon^{-2})\) time.    Randomized multilevel
Monte Carlo methods (RMLMC) that produce efficient  and unbiased estimators   of expectations of functionals  arising in SDEs are given in~\citep{GlynnRhee2015unbiased,Vihola2018}.
Exact simulation algorithms, which exist for several financial models (see~\cite[\S3]{glasserman2004Monte}), also yield unbiased estimators for prices of derivatives. More recent exact simulation methods have been developed for    Heston's stochastic volatility model~\citep{broadieKaya2006,glasserman2011gamma},  jump-diffusion processes~\citep{giesecke2013exact}, and the SABR model~\citep{cai2017exact}.  

Consider now an  Asian option with a  given maturity  monitored at \(m\) fixed dates. Even in the Black-Scholes model,  the time required to estimate the option price with variance \(O(\epsilon^{2})\)   is \(\Theta(m\epsilon^{-2})\) under the conventional Monte Carlo method, assuming the payoff variance is upper and lower bounded by constants independent of \(m\). This is because  the simulation of the underlying prices at the \(m\)  dates takes \(\Theta(m)\) time.        

This paper describes  a general multilevel framework to  price   an Asian option monitored at \(m\) dates.  The basic idea behind our approach is to (approximately) simulate the forward  prices at only a subset of the \(m\) dates at a given iteration. The forward prices at the remaining dates are then  approximated  by  the average of surrounding  forward prices. Our approach does not make any assumptions on the  nature of the stochastic process driving the underlying. It however assumes the existence of a linear relationship between  the underlying and forward prices, that the underlying price is square-integrable, and makes certain assumptions on the running time required to simulate the underlying on a discrete time grid  with a given precision. The latter condition is satisfied  in any model where the forward price process can be simulated exactly at  \(m'\)  fixed dates in \(O(m')\) expected time. Using the  Milstein scheme, it is also satisfied by processes driven by scalar SDEs. Our approach   yields  unbiased estimators  with variance \(O(\epsilon^2)\) for the Asian option price in \(O(m+\epsilon^{-2})\) expected time for a variety of processes including the Black-Scholes
model,     Merton's jump-diffusion model, the Square-Root model, Kou's
double exponential jump-diffusion model, the variance gamma and
 NIG exponential Levy processes and, using the  Milstein scheme, processes driven by scalar SDEs.   Our method is  also applicable with the same performance guarantees if the underlying is the average of  assets  that follow   a  multi-dimensional geometric Brownian motion. Using the Euler scheme, our approach  estimates the Asian option price with mean square error \(O(\epsilon^{2})\)   in   \(O(m+(\ln(\epsilon))^{2}\epsilon^{-2})\) expected time for processes driven by one-dimensional or multidimensional SDEs.
We are
 not aware of any previous Monte Carlo, MLMC or RMLMC method that provably achieves such   tradeoffs
 between the running time and  target accuracy, even under the Black-Scholes model. \citet{giles2013numerical} and \citet{giles2014antithetic} mention that their methods can be used to price Asian options monitored at \(m\) dates, but do not analyse the performance of their algorithms in terms of \(m\). Our paper makes three main contributions:\begin{enumerate}
\item 
Our approach  prices Asian options monitored at \(m\) dates and achieves a target accuracy \(O(\epsilon)\) in    \(O(m+\epsilon^{-2})\) or    \(O(m+(\ln(\epsilon))^{2}\epsilon^{-2})\) expected time, depending on the assumptions satisfied by the diffusion process.
It applies to a wide range of processes, including processes with jumps.
\item When the forward price process can be simulated exactly at  \(m'\)  fixed dates in \(O(m')\) expected time, we give explicit upper-bounds on the variance of our estimators in terms of the underlying variance at \(T\).
Certain processes such as the Square-Root diffusion satisfy this condition even though they  have no known discretization schemes with positive strong order of convergence, and so multilevel methods based on the Euler or Milstein schemes are inapplicable to such processes. 
\item  We do not make any assumptions on the dates at which the option is monitored. We  assume that the sum of the absolute values of  the weights associated with the monitoring dates is upper-bounded by a constant independent of \(m\), but make no   assumptions on the sign or order of magnitude of these weights. Our approach thus applies to average price and  average strike options.   
\end{enumerate}
The rest of the paper is organized as follows. \S\ref{se:preliminaries} describes the modelling framework and recalls the MLMC and RMLMC methods. \S\ref{se:algorithmsDescription} presents our algorithms for Asian options pricing.
Examples are described  in~\S\ref{se:examples}. Numerical simulations are given   in \S\ref{se:numericalExamples}.  We provide concluding remarks    in \S\ref{se:conclusion}. Omitted proofs are in the appendix\commentt{}{ or the electronic companion}, which also contains additional numerical results.  \section{Preliminaries}\label{se:preliminaries}
\subsection{The modelling framework}  Assume that interest rates are deterministic.  Let  \(T\) be a fixed maturity and \(m\) a positive integer. Denote by  \(F(t)\) the forward price of an underlying calculated at time \(t\) for maturity \(T\). For  \(0\le j\le m\), let   \(F_{j}=F(t_{j})\), where \(t_{0}<\cdots< t_{m}\),  with \(t_{0}=0\) and \(t_{m}=T\). Note that \(F_{m}\) is the underlying price at \(T\). Let \(A=\sum^{m}_{j=1}w_{j}F_{j}\) be a linear combination of the forward prices, where the \(w_{j}\)'s are  non-zero signed weights whose absolute values sum up to \(1\).  Consider an Asian option with payoff   \(f(A)\) at maturity \(T\), where  \(f\) is a \(\kappa\)-Lipschitz real-valued function of one variable. Such  a payoff can model  Asian options that arise in a broad range of situations. For instance, the payoff of  an average price call with strike \(K\)  and maturity \(T\) on futures prices maturing at \(T\) is equal to   \(f(A)\),  with \(f(x)=\max(x-K,0)\) and \(w_{1}=\dots=w_{m}=1/m\).  This is because forward prices are equal to futures prices when interest rates are deterministic. Similarly, the payoff of  an average strike call with maturity \(T\) on futures prices maturing at \(T\)  is  equal to   \(f(A)\), where \(f(x)=2\max(x,0)\) and  \(w_{1}=\dots=w_{m-1}=-(m-1)^{-1}/2\), with \(w_{m}=1/2\).
   In the same vein,  average price  and average strike options have a payoff  equal to \(f(A)\) for a suitable choice of \(f\) and of the weights \(w_{j}\)'s if the underlying is a stock  that pays deterministic dividends, or an index with a deterministic and continuous dividend rate, or an exchange rate. This  is due to the existence of a deterministic linear relationship between the forward price and the underlying price (see~\cite[Chap. 5]{Hull14}).  
  
We assume the existence of a risk-neutral probability \(Q\) such that the sequence \((F_{j})\), \(0\leq j\leq m\), is a martingale under \(Q\), and the price of the option at time \(0\) is 
\(e^{-rT}E(f(A))\),
 where \(r\) is the risk-free rate at time \(0\) for maturity \(T\).  The existence of \(Q\) can be shown under no-arbitrage conditions (see~\cite[\S1.2.2]{glasserman2004Monte}). All expectations in this paper are taken with respect to \(Q\). We  assume that  \(F_{m}\) is square-integrable. By~\cite[Corollary~1.6, p.~53]{Yor99}, this implies that \(F_{j}\) is  square-integrable for \(1\leq j \leq m\). We also assume that   \(\kappa\) is upper-bounded
 by a constant independent of \(m\). 
\subsection{The MLMC method}
\label{sub:MLMC}The MLMC method described in~\citep{Giles2008} efficiently estimates  the expectation of a random variable \(Y_{L}\)   that is approximated with increasing accuracy  by random variables \(Y_{l}\),  \(0\le l\leq L-1\), for some integer \(L\). For \(0\le l\leq L\), denote by   \(C_{l}\)  the expected cost of computing \(Y_{l}-Y_{l-1}\), with  \(Y_{-1}:=0\).   Assume that \(Y_{l}\),  \(0\le l\leq L\),  are   square-integrable.  For   \(0\leq l\leq L\),   let  \(\bar Y_{l}\) be the average of \(n_{l}\) independent copies of \(Y_{l}-Y_{l-1}\), where \(n_{l}\) is a positive integer to be specified later. Assume that  the estimators   \(\bar Y_{0},\dots,\bar
Y_{L}\) are independent.  Following the analysis in~\citep{Giles2008}, 
\(\bar Y=\sum^{L}_{l=0}\bar Y_{l}
\) is an unbiased estimator of \(E(Y_{L})\), and\begin{equation}\label{eq:varGiles}
\var(\bar Y)=\sum^{L}_{l=0}\frac{\mu_{l}}{n_{l}},
\end{equation}where \(\mu_{l}\triangleq\var(Y_{l}-Y_{l-1})\) for \(0\leq l\leq L\). Let  \(\bar C =\sum^{L}_{l=0}n_{l}C_{l}\) be the expected cost of computing \(\bar Y\). It is observed in~\citep{Giles2008}
that the work-normalized variance \(\bar C \var(\bar Y)\) is minimized when \(n_{l}\)        
is proportional to \(\sqrt{{\mu_{l}}/{C_{l}}}\), ignoring integrality constraints. The work-normalized variance of an unbiased estimator is defined as the  product of the variance and  expected running time. \citet{glynn1992asymptotic} show that the efficiency of an unbiased
estimator is inversely proportional to the work-normalized variance.

\subsection{The RMLMC method}We now recall a RMLMC method of~\citet{GlynnRhee2015unbiased} that efficiently estimates  the expectation of a random variable \(Y\)   that is approximated by random variables \(Y_{l}\),  \(l\geq0\). As in \S\ref{sub:MLMC}, denote by   \(C_{l}\)  the expected cost of computing \(Y_{l}-Y_{l-1}\), for \(l\geq0\),  with  \(Y_{-1}:=0\).  Assume that  \(Y\) and  \(Y_{l}\), \(l\geq0\), are   square-integrable.   Let \((p_{l})\), \(l\geq0\), be a probability distribution  such that  \(p_{l}>0\) for \(l\geq0\). Let   \(N\in\mathbb{N}\)  be an integral  random variable independent of  \((Y_{l}: l\geq0)\)  such that \(\Pr(N=l)=p_{l}\)   for \(l\geq0\). 
Set \(Z=(Y_{N}-Y_{N-1})/p_{N}\), with  \(Y_{-1}:=0\). For  a square-integrable random variable  \(X\), let \(||X||=\sqrt{E(X^{2})}\).  The following result is due to~\citet{GlynnRhee2015unbiased}  (see also~\cite[Theorem 2]{Vihola2018}).  
\begin{theorem}[\citep{GlynnRhee2015unbiased}]\label{th:Glynn}Assume that \(||Y_{l}-Y||\) converges to \(0\) as \(l\) goes to infinity.   If \(\sum^{\infty}_{l=0}||Y_{l}-Y_{l-1}||^{2}/p_{l}\) is finite  then \(Z\) is square-integrable, \(E(Z)=E(Y)\), and
\begin{equation*}
||Z||^{2}=\sum^{\infty}_{l=0}\frac{||Y_{l}-Y_{l-1}||^{2}}{p_{l}}.
\end{equation*}
\end{theorem}
Denote by   \(C\) be the expected cost of computing \(Z\). Propositions~\ref{pr:unbiasedML} and~\ref{pr:biasedML}  below are in the same spirit as results previously obtained in~\cite[Theorem 3.1]{Giles2008} and~\citep{GlynnRhee2015unbiased}. For completeness, we give their proof in the appendix. Proposition~\ref{pr:unbiasedML}  shows that, under certain conditions on \(Y_{l}\) and \(C_{l}\), the sequence \((p_{l})\), \(l\geq0\), can be chosen so that both \(||Z||\) and \(C\) are finite.
\begin{proposition}\label{pr:unbiasedML}
Assume that \(||Y_{0}||^{2}\leq\nu\) and that, for \(l\geq0\), \begin{equation}\label{eq:YlYProp}
||Y_{l}-Y||^{2}\leq\nu2^{-\beta l}
\end{equation} and    \(C_{l}\leq c2^{l}\),    where \(c\), \(\nu\) and \(\beta\) are positive constants, with \(\beta\in(1,2]\).  If, for \(l\geq0\),  
\begin{equation}\label{eq:plDefBeta}
p_{l}=(1-2^{-(\beta+1)/2}){2^{-(\beta+1)l/2}},
\end{equation}   then \(Z\) is square-integrable,  \(E(Z)=E(Y)\), and
\begin{equation}\label{eq:||Z||^2Prop}
||Z||^{2}\leq \frac{20\nu}{1-2^{-(\beta-1)/2}}.
\end{equation} 
Furthermore, \begin{equation}\label{eq:TUpperBoundProp}
C\leq\frac{c}{1-2^{-(\beta-1)/2}}.
\end{equation}
\end{proposition}
 If we relax~\eqref{eq:YlYProp}, Proposition~\ref{pr:biasedML}  shows how to construct a biased estimator \(Z_{L}\) of \(Y\), for any  positive integer \(L\), with expected cost and variance bounded by a linear function of \(L\), and a bias that decreases geometrically with \(L\).     
\begin{proposition}\label{pr:biasedML}
Assume  that \(||Y_{0}||^{2}\leq\nu\) and that, for  \(l\geq0\),  \begin{equation}\label{eq:YlYPropBias}
||Y_{l}-Y||^{2}\leq\nu2^{- l}
\end{equation} and \(C_{l}\leq c2^{l}\), where \(\nu\) and \(c\)  are positive constants.  Let \(p_{l}={2^{-(l+1)}}\) for \(l\geq0\).  Fix  a positive integer \(L\) and set  
\(Z_{L}=(Y_{N}-Y_{N-1}){\bf1}_{N\leq L}/{p_{N}}. 
\)   Then  \(Z_{L}\) is square-integrable, \begin{equation}\label{eq:SquaredBiasProp}
(E(Z_{L}-Y))^{2}\leq\nu2^{-L},
\end{equation} and
\begin{equation}\label{eq:||Z||^2PropBiased}
||Z_{L}||^{2}\leq12\nu (L+1).
\end{equation}Furthermore, the expected cost of computing \(Z_{L}\) is at most \(cL\).
\end{proposition}
More sophisticated versions of the RMLMC method can be found in~\citep{GlynnRhee2015unbiased,Vihola2018}.
\section{Multilevel algorithms for Asian options}\label{se:algorithmsDescription} We construct multilevel approximations of \(A\) in \S\ref{sub:MLAA} and use them in \S\ref{sub:exact} and~\S\ref{sub:Approx} to build   estimators of the Asian option price.  \S\ref{sub:exact}
 considers the case where forward prices can be simulated exactly, while   \S\ref{sub:Approx} treats the case where forward  prices can be simulated approximately. Set    \(a=f((\sum^{m}_{j=1}w_{j})F_{0})\) and     \(U=f(A)-a\). 
\subsection{Multilevel approximations of \(A\)}
\label{sub:MLAA}  Here we construct an increasing sequence of subsets of  \(\{1,\dots,m\}\) and show that \(A\) is approximated, with increasing accuracy, by linear combinations of  forward prices  corresponding to these subsets. For integers \(i\) and \(j\) with \(1\leq i\leq m\) and   \(0\leq j\leq m\), let \begin{displaymath}
W(i,j)=\sum_{k=i}^{j}w_{k} \text{ and }
W'(i,j)=\sum_{k=i}^{j}|w_{k}|.
\end{displaymath} By convention, \(W(i,j)=W'(i,j)=0\) if \(j<i\). Define the subsets \(J_{l}\) of \(\{1,\dots,m\},\) for \(l\geq0\), as follows. 
Set \(L=\lceil\log_{2}m\rceil\) and \(J_{l}= \{1,\dots,m\}\) for \(l\geq L\). For \(0\leq l\leq L-1\), let   \begin{equation}\label{eq:JlDef}
J_{l}=\{j\in  \{1,\dots,m\}:2^{l}W'(1,j-1)< \lfloor 2^{l}W'(1,j)\rfloor\}.
\end{equation}Note that \(J_{0}=\{m\}\). Roughly speaking, \(J_{l}\) consists of the indices \(j\) where the sequence \(W'(1,j)\) ``jumps'' over a multiple of \(2^{-l}\).  It is therefore reasonable to expect that the sequence \((J_{l})\), \(l\geq0\), is increasing and that  the size of \(J_{l}\) is at most \(2^{l}+1\).        
\begin{proposition}\label{pr:JlIncreasingandSize} 
For \(l\geq 0\), \begin{equation}\label{eq:sizeJl}
|J_{l}|\le 2^{l}+1 
\end{equation} and \begin{equation}\label{eq:subsetJl}
J_{l}\subseteq J_{l+1}.
\end{equation} 
 \end{proposition}
Proposition~\ref{pr:JlIncreasingandSize}  implies that, for  \(0\leq l\leq L-1\),  \begin{equation}\label{eq:JlBackWardInduction}
J_{l}=\{j\in  J_{l+1}:2^{l}W'(1,j-1)< \lfloor 2^{l}W'(1,j)\rfloor\}.
\end{equation} 
For \(l\geq0\), define the following trapezoidal approximation of \(A\):\begin{equation}\label{eq:AlDef}
A_{l}=\sum_{j\in J_{l}}w_{j}F_{j}+\frac{1}{2}\sum_{(i,k)\in\mathcal{P}_{l}}W(i+1,k-1)(F_{i}+F_{k}),
\end{equation}where \(\mathcal{P}_{l}\) is the set of pairs of  consecutive of elements of the set \(\{0\}\cup J_{l}\). Thus \(A_{l}\) is obtained from \(A\) by replacing each \(F_{j}\) with \((F_{i}+F_{k})/2\) for each  pair \((i,k)\in\mathcal{P}_{l}\) and each integer \(j\) with \(i<j<k\). By construction,  \(A_{l}\) is a deterministic linear function of \((F_{j})\), \(j\in J_{l}\). Note that   \(A_{l}=A\) for \(l\geq L\). Theorem~\ref{th:AsianBasicML} below gives a bound on the \(L^{2}\)-distance between \(A_{0}\) and \(W(1,m)F_{0}\) on one hand, and between \(A_{l}\) and \(A\) on the other hand. 
\begin{theorem}\label{th:AsianBasicML}\(||A_{0}-W(1,m)F_{0}||^{2}\leq\var(F_{m})\) and, for \(l\geq0\),
\begin{equation}\label{eq:MLAAl}
||A_{l}-A||^{2}\le2^{-2l}\var(F_{m}).
\end{equation}
\end{theorem}Algorithm M below calculates the coefficients \(W(i+1,k-1)\) in~\eqref{eq:AlDef}, for  \(0\leq l\leq L-1\) and \((i,k)\in\mathcal{P}_{l}\),   in   \(O(m)\)  total time,
using the following steps.
\begin{enumerate}
\item 
Calculate recursively \(W(1,j)\) and   \(W'(1,j) \) for  \(1\leq j\leq m\). \item Construct by backward induction
the subsets \(J_{l}\), for  \(0\leq l\leq L\), using~\eqref{eq:JlBackWardInduction}. This takes   \(O(m)\) total  time since \(|J_{l+1}|\le
1+2^{l+1}\) for   \(l\in\{0,\dots,L-1\}\), and so \(J_{l}\) can be constructed       in \(O(2^{l})\) time. 
\item For   \(l\in\{0,\dots,L-1\}\) and  each  pair   \((i,k)\in\mathcal{P}_{l}\),    calculate \(W(i+1,k-1)\)  via the relation \(W(i+1,k-1)=W(1,k-1)-W(1,i)\).
   For each level \(l\), this takes  \(O(2^{l})\) time, and so this step
   takes \(O(m)\) total  time.
\end{enumerate}

\subsection{The exact simulation case }
\label{sub:exact}

\begin{description}
\item[Assumption 1 (A1).] There is a constant \(c\) independent of \(m\) such that, for any subset \(J\) of \(\{1,\dots,m\}\), the expectation of the time required to simulate the vector \((F_{j})\),  \(j\in J\),  is at most \(c|J|\). 
\end{description}
A1 holds if the expectation of the time to simulate the forward price process  on a discrete time grid of size \(n\) is \(O(n)\).  
 Examples where  A1 holds are given in \S\ref{se:examples}.   Theorem~\ref{th:asian} below shows how to construct an unbiased estimator of the Asian option price under A1 using the RMLMC method.    
\begin{theorem}\label{th:asian}Suppose A1 holds. Let \(N\in\mathbb{N}\) be an integral  random variable independent of  \((F_{j}: 1\leq j\leq m)\)  such that \(\Pr(N=l)=p_{l}\)  for non-negative integer \(l\), where  \(p_{l}=(1-2^{-3/2})2^{-3 l/2}\).
Set \(V=(U_{N}-U_{N-1})/p_{N}\), where  \(U_{l}=f(A_{l})-a\) for \(l\geq0\)
and  \(U_{-1}=0\). Then \(V\) is square-integrable,
\begin{equation}\label{eq:expectedPhiExact}
E(f(A))=E(V)+a,
\end{equation}  and\begin{equation}\label{eq:||Z||^2Exact}
\var(V)\leq 70\kappa^{2}\var(F_{m}).
\end{equation} Furthermore, the expectation of the time required to simulate \(V\) is upper-bounded by a constant independent of \(m\).   
\end{theorem}
\commentt{\begin{proof}}{\proof{Proof.}}
Since \(|J_{l}|\leq2^{l}+1\), the expectation of the time  to simulate the vector (\(F_{j}\)),  \(j\in J_{l}\),  is at most \(c2^{l+1}\). Together
with~\eqref{eq:AlDef},  this implies the existence of a constant \(c'\) independent of \(m\) such that, for \(l\geq0\), the expectation of the time  to simulate \(U_{l}-U_{l-1}\) is at most \(c'2^{l}\).  Since \(A_{L}=A\), we have \(U_{L}=U\). By~\eqref{eq:AlDef},  \(A_{l}\) is square-integrable 
for \(l\geq0\) and, since \(f\) is \(\kappa\)-Lipschitz, so are \(U_{l}\) and \(U\).    As \(|U_{0}|\le \kappa|A_{0}-W(1,m)F_{0}|\), Theorem~\ref{th:AsianBasicML} implies that\begin{equation}\label{eq:||U0||^2Bound}
||U_{0}||^{2}\leq\kappa^{2}\var(F_{m}).
\end{equation} Similarly,  as \(|U_{l}-U|\le \kappa|A_l-A|\) for \(l\geq0\),
by Theorem~\ref{th:AsianBasicML},   
\begin{equation}\label{eq:UlUL2diffBound}
||U_{l}-U||^{2}\leq \kappa^{2}2^{-2l}\var(F_{m}).
\end{equation}    The conditions of  Proposition~\ref{pr:unbiasedML} are thus met for \(Y=U\) and \(Y_{l}=U_{l}\) for \(l\ge0\), with \(\nu=\kappa^2\var(F_{m})\), \(\beta=2\) and \(c=c'\). By~\eqref{eq:TUpperBoundProp},
  the expectation of the time required to simulate \(V\) is at most  \(4 c'\). Furthermore,   \(V\) is square-integrable with \(E(V)=E(U)\), which yields~\eqref{eq:expectedPhiExact}.  Similarly,~\eqref{eq:||Z||^2Exact} follows from~\eqref{eq:||Z||^2Prop}.
\commentt{\end{proof}}{\Halmos\endproof}

Theorem~\ref{th:asian} shows that
  \(e^{-rT}(V+a)\)   is an unbiased estimator of the Asian option price
  that can
  be simulated in constant time  with  variance bounded by a constant independent of \(m\). Simulating \(\lceil\epsilon^{-2}\rceil\) independent copies
  of \(V\)  
  yields an unbiased estimator  of the option price with variance \(O(\epsilon^2)\) in \(O(m+\epsilon^{-2})\) expected time,  including the \(O(m)\) preprocessing cost of
  Algorithm M.
  
  Theorem~\ref{th:MLMC} below shows how to construct another unbiased estimator of the Asian option price under A1 using the MLMC method. 

\begin{theorem}\label{th:MLMC}Suppose  A1 holds. 
Define \(U_{l}\), \(l\geq-1\), as in Theorem~\ref{th:asian} and, for \(0\leq l \leq L\), let  \(\mu_{l}=\var(U_{l}-U_{l-1})\) and \begin{equation}\label{eq:nl}
n_{l}=\big\lfloor1+ \frac{m\sqrt{{\mu_{l}}/|J_{l}|}}{\sum ^{L}_{l'=0}\sqrt{\mu_{l'}|J_{l'}|}}\big\rfloor. 
\end{equation} For   \(0\leq l\leq L\),   let  \(\bar U_{l}\) be the average of \(n_{l}\) independent copies of \(U_{l}-U_{l-1}\). Assume that  the estimators   \(\bar U_{0},\dots,\bar
U_{L}\) are independent. Set     \(\bar U=\sum^{L}_{l=0}\bar U_{l}\).  Then\begin{equation}\label{eq:expectedPhiExactMLMC}
E(f(A))=E(\bar U)+a,
\end{equation}  and\begin{equation}\label{eq:varBarU}
m\var(\bar U)\leq 240\kappa^{2}\var(F_{m}).
\end{equation} Furthermore, the expectation of the time required to simulate \(\bar U\) is  \(O(m)\).   
\end{theorem}
Assuming the variances \(\mu_{l}\), \(0\leq l \leq L\),  are known, Theorem~\ref{th:MLMC} shows that
  \(e^{-rT}(\bar U+a)\)   is an unbiased estimator of the Asian option price
  that can
  be simulated in \(O(m)\) time  with  variance \(O(1/m)\). Simulating \(\lceil\epsilon^{-2}/m\rceil\) independent copies
  of \(\bar U\)  
  yields an unbiased estimator  of the option price with variance \(O(\epsilon^2)\) in \(O(m+\epsilon^{-2})\) expected time.  The variances \(\mu_{l}\)  can be estimated by Monte
Carlo simulation.

\subsection{The approximate simulation case}
\label{sub:Approx}
For \(J\subseteq\{1,\dots,m\}\), let \(\mathbb{R}^{J}\) denote the set of vectors of dimension \(|J|\), indexed by the elements of \(J\).

\begin{description}
\item[Assumption 2 (A2).]
There are constants \(c_{1}\), \(c_{2}\) and \(\beta\in[1,2]\) such that, for  \(l\geq0\)  and  \(J\subseteq\{1,\dots,m\}\), there is a random vector  \(\hat F=\hat F(J,l)\in\mathbb{R}^{J}\) such that   \(||\hat F_{j}-F_{j}||^{2}\leq c_{2}2^{-\beta l}\) for any \(j\in J\). For \(l\geq1\) and \(J'\subseteq J\subseteq\{1,\dots,m\}\), the expected time required to simulate the vector   \((\hat F(J',l-1),\hat F(J,l))\)  is at most  \(c_{1}(|J|+2^{l})\). \end{description}
  The first condition in A2 says that, for    \(l\geq0\) and \(J\subseteq\{1,\dots,m\}\),
  the forward price  \(F_{j}\) is approximated by \(\hat F_{j}\) with  ``mean square error'' at most \(c_{2}2^{-\beta l}\) for any  \(j\in J\), where   \(\hat F=\hat F(J,l)\).  The second condition gives an upper bound on the expected time to  jointly simulate    \(\hat F(J',l-1)\) and \(\hat F(J,l)\). It is shown in \S\ref{se:EulerMil} that A2 holds under certain regularity conditions when  the Euler or Milstein schemes are used to approximately simulate  forward prices.   
\comment{
See~\citep{baran2009approximations} for processes with jumps.    
}

Assume now that A2 holds. For \(l\geq0\), let \(\hat F^{l}=\hat F(J_{l},l)\) and \begin{equation}\label{eq:hatAldef}
\hat A_{l}=\sum_{j\in J_{l}}w_{j}\hat F^{l}_{j}+\frac{1}{2}\sum_{(i,k)\in\mathcal{P}_{l}}W(i+1,k-1)(\hat F^{l}_{i}+\hat F^{l}_{k}).
\end{equation}Thus \(\hat A_{l}\) is obtained from \(A\) by replacing each \(F_{j}\) with  \(\hat F^{l}_{j}\) if \(j\in J_{l}\)  and by \((\hat F^{l}_{i}+\hat F^{l}_{k})/2\) if \((i,k)\in\mathcal{P}_{l}\) and  \(i<j<k\).   Note that \(\hat A_{l}\) is a deterministic linear function of the vector  \(\hat F^{l}\).  Proposition~\ref{pr:Approx} below gives a bound on the \(L^{2}\)-distance between \(\hat A_{0}\) and \(W(1,m)F_{0}\) on one hand, and between \(\hat A_{l}\) and \(A\) on the other hand.
\begin{proposition}\label{pr:Approx}
If A2 holds then    \(||\hat A_{0}-W(1,m)F_{0}||^{2}\leq c_{3}\) and \(||\hat A_{l}-A||^{2}\leq c_{3}2^{-\beta l}\) for \(l\geq0\), where \(c_{3}=2(c_{2}+\var(F_{m}))\). \end{proposition}

Theorem~\ref{th:asianApprox}  below shows how to construct an unbiased estimator of the Asian option price under A2, with   \(\beta>1\). The case \(\beta=1\) will be  considered in Theorem~\ref{th:asianApproxBis}.   
\begin{theorem}\label{th:asianApprox}Suppose A2 holds with \(\beta>1\). Let \(N\in\mathbb{N}\) be an integral random variable independent of  \((\hat F(J_{l},l): l\geq 0)\)  such that \(\Pr(N=l)=p_{l}\) for non-negative integer \(l\), where \(p_{l}\) is given by~\eqref{eq:plDefBeta}. Let \(\hat U_{l}=f(\hat A_{l})-a\) for \(l\geq0\), and let  
\(
\hat V=(\hat U_{N}-\hat U_{N-1})/{p_{N}}
\), where \(\hat U_{-1}:=0\). Then \(\hat V\) is square-integrable and\begin{equation}\label{eq:expectedPhiApprox}
E(f(A))=E(\hat V)+a.
\end{equation}  Furthermore, \(\var(\hat V)\) and the expectation of the time required to simulate \(\hat V\) are upper-bounded by constants independent of \(m\). \end{theorem}
As per the discussion following
Theorem~\ref{th:asian}, Theorem~\ref{th:asianApprox} shows that
  \(e^{-rT}(\hat V+a)\)   is an unbiased estimator of the Asian option price that can
  be simulated in constant time
  and with  variance bounded by a constant independent of \(m\). Independent
   \(\lceil\epsilon^{-2}\rceil\) runs of this estimator yield an unbiased estimator  of the Asian option price with variance \(O(\epsilon^2)\) in \(O(m+\epsilon^{-2})\) expected time. 

Theorem~\ref{th:asianApproxBis} below constructs an estimator of the option price with an arbitrarily small bias when A2 holds with \(\beta=1\).
\begin{theorem}\label{th:asianApproxBis}Suppose A2 holds with \(\beta=1\). Fix \(\epsilon\in(0,1/2)\) and set \(L=\lceil2\log_{2}(1/\epsilon)\rceil\). Let \(N\in\mathbb{N}\) be an integral  random variable independent of  \((\hat F(J_{l},l): l\geq 0)\)  such that \(\Pr(N=l)=2^{-(l+1)}\) for  \(l\in\mathbb{N}\). Let \(\hat U_{l}=f(\hat A_{l})-a \) for \(l\geq0\), and let   
\begin{equation*}
\hat V=\frac{\hat U_{N}-\hat U_{N-1}}{p_{N}}{\bf1}_{N\leq L},
\end{equation*}where \(\hat U_{-1}:=0\). Then \(\hat V\) is square-integrable and\begin{equation}\label{eq:expectedPhiBias}
(E(\hat V)+a-E(f(A)))^{2}\leq c_{3}\kappa^{2}\epsilon^{2},
\end{equation}  where \(c_{3}\) is defined as in Proposition~\ref{pr:Approx}. Furthermore, there are constants \(c_{4}\) and \(c_{5}\) independent of \(m\) and of \(\epsilon\) such that \(\var(\hat V)\leq c_{4}\ln(1/\epsilon)\) and the expectation of the time required to simulate \(\hat V\) is upper-bounded by \( c_{5}\ln(1/\epsilon)\). 
\end{theorem}
Under the assumptions of Theorem~\ref{th:asianApproxBis}, the Asian option price  can be calculated with \(O(\epsilon^{2})\) mean square error in   \(O(m+\epsilon^{-2}\ln^{2}(1/\epsilon))\) expected time as follows. We simulate  \(n\) independent copies of \(\hat V\), where \(n=\lceil\ln(1/\epsilon)\epsilon^{-2}\rceil\), and take their average \(\hat V_{n}\). Since \(\var(\hat V_{n})=\var(\hat V)/n\), we have    \(\var(\hat V_{n})\le c_{4}\epsilon^{2}\).  Furthermore,
as \(E(\hat V_{n})=E(\hat V)\), it follows from~\eqref{eq:expectedPhiBias} that \begin{displaymath}
(E(\hat V_{n})+a- E(f(A)))^{2}\leq c_{3}\kappa^{2}\epsilon^{2}.
\end{displaymath}Since the mean square error is the sum of   the variance
 and squared bias, we conclude that\begin{displaymath}
||\hat V_{n}+a- E(f(A))||^{2}\leq (c_{4}+c_{3}\kappa^{2})\epsilon^{2}.
\end{displaymath}Thus \(e^{-rT}(\bar V+a)\) is an estimate of the Asian option
price \(e^{-rT} E(f(A))\) with mean square   error \(O(\epsilon^{2})\). The total expected time  to simulate  \(\hat V_{n}\)  is  \(O(m+\ln^{2}(\epsilon)\epsilon^{-2})\), including the cost of Algorithm M.
\section{Examples}
\label{se:examples}
 Below are examples where  A1 holds. 
\subsection{The Black-Scholes model}\label{sub:exBS}In this model,  \(F(t)\)  satisfies the SDE  \begin{equation}\label{eq:BS}
d F(t)=\sigma F(t)dW
\end{equation}   on \([0,T]\), where  \(\sigma\) is a  constant volatility and   \(W\) is a one-dimensional Brownian motion under \(Q\).
Given \(J\subseteq\{1,\dots,m\}\), let \(n=|J|\), and let \(0=\tau_{0}<\tau_{1}<\cdots<\tau_{n}\) be the elements of the time grid \(G=\{0\}\cup \{t_{j}:j\in J\}
\), sorted in increasing order.  Let \(X_{1},\dots,X_{n}\) be independent standard Gaussian random variables. We simulate the forward prices on  \(G\) in \(O(n)\) time using the following recursive procedure~\cite[\S3.2.1]{glasserman2004Monte}:\begin{equation*}
F(\tau_{k})= F(\tau_{k-1})\exp(-\sigma^{2}\frac{\tau_{k}-\tau_{k-1}}{2}+\sigma\sqrt{\tau_{k}-\tau_{k-1}}X_{k}),
\end{equation*}    \(1\leq k\leq n\). Then, for \(j\in J\), we set \(F_{j}=F(\tau_{k})\). where \(k\) is the index such that \(\tau_{k}=t_{j}\).
Thus  A1 holds for the Black-Scholes
model. Furthermore, it is well-known that the forward price is square-integrable at any fixed date in this model.
\subsection{Merton's jump-diffusion model}
\label{sub:exMerton}
The risk-neutral process for the forward price in this  model (see~\citep{merton1976option}) is:\begin{equation*}
\frac{dF(t)}{F(t-)}=-\lambda mdt+\sigma dW(t)+
dJ(t)
\end{equation*}
 on \([0,T]\), where \(W\) is a Brownian motion,
\(J(t)=\sum_{j=1}^{N(t)}(Y_j-1)\),
and
\(N(t)\)
is a Poisson process with rate \(\lambda\). If  a jump occurs at time
\(\tau_{j}\), then \(S(\tau_{j}+)=S(\tau_{j}-)Y_{j}\),
 where \(\ln(Y_{j})\) is a Gaussian random variable with mean \(\beta\) and standard
 deviation \(\gamma\).  The  model parameters satisfy the equation:
  \(m+1=\exp(\beta+\gamma^{2}/2)\). We assume that \(W\), \(N\) and the  \(Y_{j}\)'s
 are independent. An algorithm that simulates the forward price process on a discrete time grid of size \(n\) in \(O(n)\) expected time  is given in~\cite[\S3.5.1]{glasserman2004Monte}. Thus  A1 holds for Merton's jump-diffusion model.
A classical calculation based on~\cite[\S3.5.1]{glasserman2004Monte}  shows that  the forward price is square-integrable at any fixed date in this model. 
\subsection{ The Square-Root diffusion model}
Here we assume that   \(F(t)\) satisfies the following SDE: 
\begin{equation*}
dF(t)=\sigma \sqrt{F(t)} dW(t)
\end{equation*}
 on \([0,T]\), where \(W\) is a Brownian motion under \(Q\), and \(\sigma>0\).
The  Square-Root diffusion model, introduced in~\citep{cox1976valuation}, is a special case of the CEV model.  An algorithm that simulates the forward price process on a discrete time grid of size \(n\) in \(O(n)\) expected time is described in \S\ref{se:simulationOfBESQ0}.    Thus  A1 holds for the Square-Root diffusion model. It is also shown in \S\ref{se:simulationOfBESQ0} that \(F_{m}\) is square-integrable.  

  It is well-known that the standard Euler scheme is not defined for Square-Root diffusions because it may produce negative forward prices. The related Cox-Ingersoll-Ross process has an implicit Euler scheme with  a strong convergence of order $1$ (see~\citep[\S3.2]{alfonsi2015affine})   under
certain assumptions on the model parameters, but we are not aware of  discretization schemes with  positive strong order of convergence for Square-Root diffusions.
\subsection{Other examples} 
It can be shown that A1 holds for a variety of other processes such as Kou's
double exponential jump-diffusion model (see~\citep{Kou2002Jump}), and   the variance gamma and
 NIG  exponential Levy processes. Algorithms that simulate these processes on a discrete time-grid are described in~\cite[\S3.5]{glasserman2004Monte}, and  it is easy to prove that the underlying second moment is finite under certain conditions on the model parameters. A1 also holds if the underlying is the average of  assets  that follow   a  multi-dimensional geometric Brownian motion. An algorithm that jointly simulates such assets is given in~\cite[\S3.2.3]{glasserman2004Monte}.  \comment{ The CEV model is studied thoroughly in~\citep{davydov2001pricing,jeanblanc2009mathematical}.
}\section{Numerical experiments}\label{se:numericalExamples}
  We have implemented the RMLMC method  of Theorem~\ref{th:asian}, and the MLMC method   of Theorem~\ref{th:MLMC}, but replaced  \(m\)  with \(30m\) in~\eqref{eq:nl} in order to mitigate the rounding
effect and achieve greater efficiency. The variances \(\mu_{l}\) were estimated by Monte Carlo simulation using \(10^{4}\) independent runs.  The  RMLMC method based on the Milstein scheme  (RMLMC-Milstein) was implemented for the Black-Scholes model as described in Theorem~\ref{th:asianApprox}, with \(\beta=2\),  without solving explicitly~\eqref{eq:BS}. The codes 
were written  in the C++ programming language. Our
experiments assume that interest rates are constant and equal to \(r\). In Tables~\ref{tab:BS} through~\ref{tab:SQRAvgPriceStrike},   ``Price'' is the estimated Asian option price obtained via  \(n\) independent replications, and ``Std'' is the estimated price   standard error. The variable    ``Cost'' refers to the total number of simulated underlying  prices throughout the \(n\) replications.  Thus, \(\text{Cost}\times\text{Std}^2\) is an estimate of the work-normalized variance. In each table, the number of independent replications is chosen so that the variable    ``Cost'' has the same order of magnitude for the studied algorithms. As the variance of a single run of the standard Monte Carlo estimator is \(e^{-2rT}\var(f(A))\), the  variance reduction factor VRF is defined as\begin{displaymath}
\text{VRF}=\frac{ 
me^{-2rT}\var(f(A))
 }{\text{Cost}\times\text{Std}^2},
\end{displaymath}
where  \(\var(f(A))\) is estimated via $10^5$ independent samples of \(A\).  The payoff of an average price call with strike \(K\) is \(\max(m^{-1}(\sum^{m}_{i=1}S_{i})-K,0)\), while the payoff of an average strike call is \(\max(S_{m}-(m-1)^{-1}(\sum^{m-1}_{i=1}S_{i}),0)\), where \(S_{i}\) is the underlying price at \(t_{i}=iT/m\).     
  
\subsection{The Black-Scholes model}
In our experiments,  the underlying is a stock \(S\) with no dividends, and
the model parameters are \(S_0=2\), \(\sigma=50\%\), 
 \(r=5\%\), and \(T=2\). These values are taken from~\citep{LinetskyAsian2004}. Table~\ref{tab:BS} gives our results for average price calls with \(K=2\) and selected values of \(m\).    The cost of a single replication, i.e. \(\text{Cost}/n\), is roughly independent of \(m\) for  the RMLMC and RMLMC-Milstein algorithms, and is roughly proportional to \(m\) for the MLMC  algorithm. For the RMLMC, MLMC and RMLMC-Milstein algorithms, the products  \(\text{Cost}\times\text{Std}^2\) are roughly independent of \(m\), and the VRFs are roughly proportional to \(m\). These results are consistent with  Theorems~\ref{th:asian},~\ref{th:MLMC} and~\ref{th:asianApprox}. Table~\ref{tab:BSAvgStrike} reports similar results for average strike calls. In Table~\ref{tab:BS},   the RMLMC and MLMC methods have a similar performance, and  slightly outperform the  RMLMC-Milstein algorithm. In Table~\ref{tab:BSAvgStrike},  the MLMC method  slightly outperforms the RMLMC method. This can be explained by observing that the  frequencies \(n_{l}\) in Theorem~\ref{th:MLMC} are near-optimal, which is not always the case for the probabilities \(p_{l}\) in Theorem~\ref{th:asian}. The RMLMC method outperforms the  RMLMC-Milstein algorithm by about a factor of \(2\). In practice, the price of a continuously monitored Asian option   can be approximated   by using a  very large value of \(m\), as reported in Table~\ref{tab:BSAvgPriceStrike}. The  price of the average price call   produced by the  RMLMC algorithm in Table~\ref{tab:BSAvgPriceStrike} is very close
to the price of the  continuously monitored average price  call given in~\citep{LinetskyAsian2004},
which is \(0.350095\).    

\subsection{Merton's jump-diffusion model}
\label{sub:Merton}

In our 
experiments,  the underlying is an index with constant dividend yield \(q\).
The model parameter values used  are  \(S_0=2\),  \(\sigma=17.65\%\),   $r = 5.59\%$, $q = 1.14\%$, \(\lambda= 8.90\%\), \(\beta=-88.98\%\), and \(\gamma=45.05\%\).
Except for the spot price, these values are taken from~\citep{AA00},
 where they were obtained by fitting option prices with maturities ranging
 from one month  to ten years.
 We  set \(T=2\). Tables~\ref{tab:MJD} and~\ref{tab:MJDAvgStrike} give prices of average price and average strike calls, respectively, using the RMLMC and MLMC algorithms. The estimated work-normalized variances of the RMLMC and MLMC methods are roughly independent of \(m\), and the VRFs are roughly proportional to \(m\). The RMLMC and MLMC methods have a similar performance for average price  calls, but  MLMC   slightly   outperforms RMLMC for  average strike calls.

\subsection{The Square-Root diffusion model}
The model parameter values in our  experiments are \(S_{0}=2\), \(r=5\%\), \(\sigma=0.4\), and \(T=2\).
Tables~\ref{tab:SQR} and~\ref{tab:SQRAvgStrike} give prices of average price and average strike calls, respectively, using the RMLMC and MLMC algorithms. Our simulation results are similar in nature to those of the Black-Scholes model and Merton's jump-diffusion model.

  \section{Conclusion}
\label{se:conclusion}
We have described  a general MLMC framework to estimate the price of an Asian option  monitored at \(m\) dates. We assume the existence of a linear relation between the underlying and forward prices, and that the underlying price
is square-integrable at maturity \(T\). Our approach   yields   unbiased estimators  with variance \(O(\epsilon^2)\) in \(O(m+\epsilon^{-2})\) expected time for a variety of processes that can be simulated exactly and,    via the Milstein scheme, processes driven by scalar SDEs.  Using the Euler scheme,  our approach  estimates the Asian option price with mean square error \(O(\epsilon^{2})\)   in   \(O(m+(\ln(\epsilon))^{2}\epsilon^{-2})\) expected time for processes driven by multidimensional SDEs.
 Numerical experiments  confirm that our approach outperforms the conventional Monte Carlo method by a factor of order \(m\).
\swap{
\commentt{\appendix}
{
\begin{APPENDIX}{}
}
\section{Proof of Proposition~\ref{pr:unbiasedML}}
Since \((x+x')^{2}\leq2(x^{2}+x'^{2})\) for any real numbers \(x\) and \(x'\), if \(X\) and \(X'\) are square-integrable random variables,\begin{equation}\label{eq:CauchyXX'}
||X+X'||^{2}\leq2(||X||^{2}+||X'||^{2}).
\end{equation} For \(l\geq1\), by applying~\eqref{eq:CauchyXX'} with \(X=Y_{l}-Y\) and \(X'=Y_{l-1}-Y\), it follows that
\begin{equation}\label{eq:YlYlm1GenDiff}
||Y_{l}-Y_{l-1}||^{2}\leq2(||Y_{l}-Y||^{2}+||Y_{l-1}-Y||^{2}).
\end{equation}Since \(||Y_{l-1}-Y||^{2}\leq4\nu2^{-\beta l}\) by~\eqref{eq:YlYProp}, it follows that from~\eqref{eq:YlYlm1GenDiff} that\begin{equation}\label{eq:YlYlm1Prop}
||Y_{l}-Y_{l-1}||^{2}\le10\nu2^{-\beta l}.
\end{equation} As \(||Y_{0}||^{2}\leq\nu\),~\eqref{eq:YlYlm1Prop} holds also for \(l=0\). Thus, as  \(p_{l}\ge 2^{-1-(\beta +1)l/2}\),\begin{eqnarray*}\sum^{\infty}_{l=0}\frac{||Y_{l}-Y_{l-1}||^{2}}{p_{l}}&\le&20\nu\sum^{\infty}_{l=0}2^{-(\beta-1)l/2}\\
&=&\frac{20\nu}{1-2^{-(\beta-1)/2}}.
\end{eqnarray*}  By  Theorem~\ref{th:Glynn}, we conclude that \(Z\) is square-integrable with \(E(Z)=E(Y)\), and that~\eqref{eq:||Z||^2Prop} holds.

We now prove~\eqref{eq:TUpperBoundProp}. As observed in~\citep{GlynnRhee2015unbiased},  \(C=\sum^{\infty}_{l=0}p_{l}C_{l}\). 
Since \(p_{l}\le 2^{-(\beta +1)l/2}\),
\begin{equation*}
C\le c\sum^{\infty}_{l=0}2^{-(\beta -1) l/2},
\end{equation*} which concludes the proof.   \qed
\section{Proof of Proposition~\ref{pr:biasedML}}
   We  apply Theorem~\ref{th:Glynn} to the sequence \((Y_{\min(l,L)}:l\geq0)\)
   and \(Y_{L}\). Thus \(Z=Z_{L}\), and so \(Z_{L}\) is square-integrable, \(E(Z_{L})=E(Y_{L})\), and 
\begin{equation*}
||Z_{L}||^{2}=\sum^{L}_{l=0}\frac{||Y_{l}-Y_{l-1}||^{2}}{p_{l}}.
\end{equation*} Hence
\begin{eqnarray*}
(E(Z_{L}-Y))^{2}&=&(E(Y_{L}-Y))^{2}\\
&\le&||Y_{L}-Y||^{2},
\end{eqnarray*}
which yields~\eqref{eq:SquaredBiasProp}. On the other hand, for \(l\geq1\), as \(||Y_{l-1}-Y||^{2}\leq2\nu2^{-l}\) by~\eqref{eq:YlYPropBias}, it follows from~\eqref{eq:YlYlm1GenDiff} that \begin{equation}\label{eq:YlYlm1Prop1}
||Y_{l}-Y_{l-1}||^{2}\le6\nu2^{-l}.
\end{equation}Since \(||Y_{0}||^{2}\leq\nu\),~\eqref{eq:YlYlm1Prop1} also holds for \(l=0\). Hence, \begin{equation*}\sum^{L}_{l=0}\frac{||Y_{l}-Y_{l-1}||^{2}}{p_{l}}\le12\nu(L+1),
\end{equation*}  which implies~\eqref{eq:||Z||^2PropBiased}. Finally, the  expected cost of computing \(Z_{L}\) is   \(\sum^{L}_{l=0}p_{l}C_{l}\), which is   upper-bounded by \(cL\)  since \(p_{l}C_{l}\le c/2\).
\qed

\section{Proof of Proposition~\ref{pr:JlIncreasingandSize}  }
We first show~\eqref{eq:sizeJl}. As this equation clearly holds for \(l\geq L\), we assume that \(0\leq l\leq L-1\).  Let \(j\) and \(j'\) be two elements of \(J_{l}\), with \(j<j'\). As \(j\leq j'-1\), 
\begin{eqnarray*}\lfloor2^{l}W'(1,j)\rfloor
&\le&\lfloor2^{l}W'(1,j'-1)\rfloor\\
&\le&2^{l}W'(1,j'-1)\\&<&\lfloor2^{l}W'(1,j')\rfloor,\end{eqnarray*}  where the last equation follows from~\eqref{eq:JlDef}. Thus the map \(j\mapsto \lfloor 2^{l}W'(1,j)\rfloor\) from \(J_{l}\) to \(\{0,\dots,2^{l}\}\) is strictly increasing. This implies~\eqref{eq:sizeJl}. 

We now show~\eqref{eq:subsetJl}.  As this relation is obvious when \(l\ge L-1\),  assume that   \(0\leq l\leq L-2\). Since \(2\lfloor x\rfloor\le\lfloor2 x\rfloor\) for \(x\in\mathbb{R}\), for any  an element \(j\) of \(J_{l}\),\begin{equation*} 2^{l+1}W'(1,j-1)
<2\lfloor 2^{l}W'(1,j)\rfloor\le\lfloor 2^{l+1}W'(1,j)\rfloor,\end{equation*}  where the first equation follows from~\eqref{eq:JlDef}. Thus, \(j\in J_{l+1}\).  This implies~\eqref{eq:subsetJl}.
\qed
\section{Proof of Theorem~\ref{th:AsianBasicML}}

 Proposition~\ref{pr:martingale1} below proves standard properties of square-integrable
 martingales.
 \begin{proposition}\label{pr:martingale1} For  \(0\leq i\leq j\leq k\le m\), \begin{equation}\label{eq:martijk}
E(F_{i}(F_{k}-F_{j}))=0,
\end{equation}and  \begin{equation}\label{eq:convMartingale}
||F_{i}||\leq||F_{j}||.
\end{equation} Moreover, \begin{equation}\label{eq:martingaleIneqBis}
||F_{j}-F_{i}||^{2}\le ||F_{k}||^{2}-||F_{i}||^{2}.
\end{equation} \end{proposition}\commentt{\begin{proof}}{\proof{Proof.}}Let \(\mathcal{F}=(\mathcal{F}_{i})\), \(0\leq i\leq m\), be the natural filtration of the random process  \(({F}_{i})\), \(0\leq i\leq m\). By the tower law,  \begin{eqnarray*}
E(F_{i}(F_{k}-F_{j}))
&=& E(E(F_{i}(F_{k}-F_{j})|\mathcal{F}_{j}))\\
&=& E(F_{i}E(F_{k}-F_{j}|\mathcal{F}_{j}))
\\
&=&0.
\end{eqnarray*}
The last equation follows from the fact that   \(({F}_{i})\), \(0\leq i\leq m\), is a martingale with respect to \(\mathcal{F}\).
This implies~\eqref{eq:martijk}. In particular, \(E(F_{i}(F_{j}-F_{i}))=0 \). As \(F_{j}=(F_{j}-F_{i})+F_{i}\),  \begin{displaymath}
||F_{j}||^{2}=||F_{j}-F_{i}||^{2}+||F_{i}||^{2},
\end{displaymath} which proves~\eqref{eq:convMartingale}. The inequality \(||F_{j}||\leq||F_{k}||\) then implies~\eqref{eq:martingaleIneqBis}.  
\commentt{\end{proof}}{\Halmos\endproof}
We next prove the following proposition.
\begin{proposition}\label{pr:WJl} For  \(l\ge0\), if \((i,k)\in\mathcal{P}_{l}\)
then \(W'(i+1,k-1)\leq2^{-l}\).
\end{proposition}
\commentt{\begin{proof}}{\proof{Proof.}}
The desired inequality clearly holds if \(k=i+1\). Assume that \(k>i+1\). Thus \(l\leq L-1\). For any integer \(j\) in \([i+1,k-1]\),  since \(j\notin J_{l}\), we have \(2^{l}W'(1,j-1)\ge\lfloor  2^{l}W'(1,j)\rfloor\), and so
\begin{equation}\label{eq:constWeights}
\lfloor2^{l}W'(1,j-1)\rfloor= \lfloor 2^{l}W'(1,j)\rfloor.
\end{equation}  
Hence 
\begin{eqnarray*}2^{l}W'(1,k-1)-1
&\le&\lfloor 2^{l}W'(1,k-1)\rfloor\\&=&\lfloor 2^{l}W'(1,i)\rfloor\\
&\le& 2^{l}W'(1,i).\end{eqnarray*}The second equation follows from~\eqref{eq:constWeights}. As \(W'(i+1,k-1)=\) \(W'(1,k-1)-W'(1,i)\), this completes the proof.
\commentt{\end{proof}}{\Halmos\endproof}
We now prove Theorem~\ref{th:AsianBasicML}. By~\eqref{eq:AlDef} and the relation \(J_{0}=\{m\}\),\begin{displaymath}
A_{0}=w_{m}F_{m}+\frac{1}{2}W(1,m-1)(F_{0}+F_{m}).
\end{displaymath}As \(W(1,m)=W(1,m-1)+w_{m}\), it follows that\begin{displaymath}
A_{0}-W(1,m)F_{0}=(\frac{1}{2}W(1,m-1)+w_{m})(F_{m}-F_{0}),
\end{displaymath} and so  \(||A_{0}-W(1,m)F_{0}||\leq||F_{m}-F_{0}||\). As \(E(F_{m})=F_{0}\), this implies the desired bound on \(||A_{0}-W(1,m)F_{0}||^{2}\). 

Fix now \(l\geq0\). For \((i,k)\in\mathcal{P}_{l}\), let \begin{displaymath}
B_{i}=\sum_{j=i+1}^{k-1}w_{j}(F_{j}-F_{i})
\text{ and } 
B'_{i}=\sum_{j=i+1}^{k-1}w_{j}(F_{j}-F_{k}).
\end{displaymath}Rewriting~\eqref{eq:AlDef} as\begin{equation*}
A_{l}=\sum_{j\in J_{l}}w_{j}F_{j}+\frac{1}{2}\sum_{(i,k)\in\mathcal{P}_{l}}\sum_{j=i+1}^{k-1}w_{j}(F_{i}+F_{k}),
\end{equation*}and noting that \begin{equation*}
A=\sum_{j\in J_{l}}w_{j}F_{j}+\sum_{(i,k)\in\mathcal{P}_{l}}\sum_{j=i+1}^{k-1}w_{j}F_{j},
\end{equation*} it follows that \begin{equation*}
A-A_{l}=\frac{1}{2}\sum_{(i,k)\in\mathcal{P}_{l}}(B_{i}+B'_{i}).
\end{equation*}
Hence, by the triangular inequality,
\begin{equation}\label{eq:AlBoundNorm}
||A-A_{l}||\leq\frac{1}{2}||\sum_{(i,k)\in\mathcal{P}_{l}}B_{i}||+\frac{1}{2}||\sum_{(i,k)\in\mathcal{P}_{l}}B'_{i}||.
\end{equation} We bound each of the two terms in the RHS of~\eqref{eq:AlBoundNorm} separately. First observe that if \((i,k)\) and \((i',k')\) are two distinct elements of \(\mathcal{P}_{l}\) with \(i<i'\), then\begin{eqnarray*}
E(B_{i}B_{i'})&=&\sum_{j=i+1}^{k-1}\sum_{j'=i'+1}^{k'-1}w_{j}w_{j'}E((F_{j}-F_{i})(F_{j'}-F_{i'}))
\\&=&0,
\end{eqnarray*}where the second equation follows from~\eqref{eq:martijk}. Thus   \begin{equation*}
||\sum_{(i,k)\in\mathcal{P}_{l}}B_{i}||^{2}=\sum_{(i,k)\in\mathcal{P}_{l}}||B_{i}||^{2}.
\end{equation*} On the other hand, for \((i,k)\in\mathcal{P}_{l}\), by the triangular inequality, 
\begin{eqnarray*}
||B_{i}||&\le&\sum_{j=i+1}^{k-1}|w_{j}|\,||F_{j}-F_{i}||^{}\\
&\le& W'(i+1,k-1)\sqrt{||F_{k}||^{2}-||F_{i}||^{2}},
\end{eqnarray*}where the second equation follows from~\eqref{eq:martingaleIneqBis}.
Using Proposition~\ref{pr:WJl}, it follows that \begin{eqnarray*}\sum_{(i,k)\in\mathcal{P}_{l}}||B_{i}||^{2} &\le& 2^{-2l}\sum_{(i,k)\in\mathcal{P}_{l}}(||F_{k}||^{2}-||F_{i}||^{2})
\\&=&2^{-2l}(||F_{m}||^{2}-||F_{0}||^{2})
\\&=&2^{-2l}\var(F_{m}).
\end{eqnarray*}
We conclude that\begin{displaymath}
||\sum_{(i,k)\in\mathcal{P}_{l}}B_{i}||\leq2^{-l}\std(F_{m}).
\end{displaymath} 
The same upper bound  on \(||\sum_{(i,k)\in\mathcal{P}_{l}}B'_{i}||\)  can be shown in a similar way.
Hence \begin{displaymath}
||A-A_{l}||\leq2^{-l}\std(F_{m}).
\end{displaymath}This concludes the proof.\qed
\section{Proof of Theorem~\ref{th:MLMC}}
The proof is similar to that of Theorem~\ref{th:asian}. As  \(U_{L}=U\), the analysis of \S\ref{sub:MLMC}, with \(Y_{l}=U_{l}\) for \(0\leq l\leq L\), shows that  \(E(\bar U)=E(U)=E(f(A))-a\). This implies~\eqref{eq:expectedPhiExactMLMC}. Let \begin{displaymath}
\bar m= \frac{m}{\sum ^{L}_{l=0}\sqrt{\mu_{l}|J_{l}|}}.
\end{displaymath} Since \(n_{l}\ge \bar m\sqrt{{\mu_{l}}/{|J_{l}|}}\) for \(0\leq l\leq L\),  it follows from~\eqref{eq:varGiles} that
\begin{eqnarray}\label{eq:varBarrU}\nonumber
\var(\bar U)&\le& \bar m^{-1}(\sum ^{L}_{l=0}\sqrt{\mu_{l}|J_{l}}|)\\
&=&\frac{(\sum ^{L}_{l=0}\sqrt{\mu_{l}|J_{l}}|)^2}{m}.
\end{eqnarray}  
As \(\mu_{0}=\var(U_{0})\), by~\eqref{eq:||U0||^2Bound}, we have \(\mu_{0}\leq\kappa^{2}\var(F_{m})\). By arguments similar to those leading to~\eqref{eq:YlYlm1GenDiff}, for \(l\ge1\),
\begin{equation*}
||U_{l}-U_{l-1}||^{2}\leq2(||U_{l}-U||^{2}+||U_{l-1}-U||^{2}).
\end{equation*} 
Since \(||U_{l-1}-U||^{2}\leq4\kappa^{2}2^{-2l}\var(F_{m})\) by~\eqref{eq:UlUL2diffBound}, 
\begin{equation*}
||U_{l}-U_{l-1}||^{2}\le10\kappa^{2}2^{-2l}\var(F_{m}).
\end{equation*} We conclude that \(\mu_{l}\leq10\kappa^{2}2^{-2l}\var(F_{m})\) for \(0\le l\leq L\).
 Since \(|J_{l}|\leq2^{l+1}\), it follows from~\eqref{eq:varBarrU} that
\begin{displaymath}
m\var(\bar U)\le\frac{20\kappa^{2}\var(F_{m})}{(1-2^{-1/2})^{2}},
\end{displaymath}
which implies~\eqref{eq:varBarU}. 

Denote by  \(C_{l}\) is the expectation of the time  to simulate \(U_{l}-U_{l-1}\), for \(0\leq l\leq L\), and let  \(\bar C =\sum^{L}_{l=0}n_{l}C_{l}\) be the expected cost of computing \(\bar Y\).  As in the proof of Theorem~\ref{th:asian}, it can be shown that there is a constant \(c'\) independent of \(m\) such that \(C_{l}\leq c'|J_{l}|\) for \(0\leq l\leq L\). As \(n_{l}\leq1+\bar m\sqrt{{\mu_{l}}/{|J_{l}|}}\),  \begin{displaymath}
\bar C \le c'\sum ^{L}_{l=0}|J_{l}|+c'\bar m\sum^{L}_{l=0}\sqrt{\mu_{l}|J_{l}|}. \end{displaymath}Since \(|J_{l}|\leq 2^{l+1}\) for \(l\geq0\), it follows that \(\bar C\leq c'2^{L+2}+c'm\leq 9c'm\).    
\qed 
\section{Proof of Proposition~\ref{pr:Approx}}
 By~\eqref{eq:AlDef} and~\eqref{eq:hatAldef}, 
 \begin{equation*}
\hat A_{l}-A_{l}=\sum_{j\in J_{l}}w_{j}(\hat F^l_{j}-F_{j})+\frac{1}{2}\sum_{(i,k)\in\mathcal{P}_{l}}W(i+1,k-1)((\hat F^l_{i}-F_{i})+(\hat F^l_{k}-F_{k})).
\end{equation*}
Hence 
 \begin{equation*}
||\hat A_{l}-A_{l}||\leq\sum_{j\in J_{l}}w_{j}||\hat F^l_{j}-F_{j}||+\frac{1}{2}\sum_{(i,k)\in\mathcal{P}_{l}}W(i+1,k-1)(||\hat F^l_{i}-F_{i}||+||\hat F^l_{k}-F_{k}||).
\end{equation*}As \(||\hat F^l_{j}-F_{j}||\leq\sqrt{c_{2}2^{-\beta l}}\) for \(j\in J_{l}\) and\begin{displaymath}
\sum_{j\in J_{l}}w_{j}+\sum_{(i,k)\in\mathcal{P}_{l}}W(i+1,k-1)=1,
\end{displaymath} it follows that
\(||\hat A_{l}-A_{l}||\leq\sqrt{c_{2}2^{-\beta l}}\). 
 Together with~\eqref{eq:MLAAl} and~\eqref{eq:CauchyXX'}, this shows that  \(||\hat A_{l}-A||^{2}\leq c_{3}2^{-\beta l}\). Similarly, as \(|| A_{0}-W(1,m)F_{0}||^{2}\leq \var(F_{m})\), we have   \(||\hat A_{0}-W(1,m)F_{0}||^{2}\leq c_{3}\).
\qed
\section{Proof of Theorem~\ref{th:asianApprox}}
The proof is similar to that of  Theorem~\ref{th:asian}. By A2 and~\eqref{eq:sizeJl}, the vector \((\hat F^{l-1},\hat F^{l})\) can be simulated in \(O(2^{l})\) expected time for \(l\geq1\). Hence, by~\eqref{eq:hatAldef}, there is a  constant   \(c'\) independent of \(m\)  such that, for \(l\geq0\),  the expectation of the time  to simulate \(\hat U_{l}-\hat U_{l-1}\) is at most \(c'2^{l}\).     As \(|\hat U_{0}|\le \kappa|\hat A_{0}-W(1,m)F_{0}|\), Proposition~\ref{pr:Approx} implies that \(||\hat U_{0}||^{2}\leq c_{3}\kappa^{2}\), where \(c_{3}\) is defined as in Proposition~\ref{pr:Approx}. Similarly, for \(l\geq0\), as \(|\hat U_{l}-U|\le \kappa|\hat A_l-A|\), Proposition~\ref{pr:Approx}  shows that 
\(||\hat U_{l}-U||^{2}\leq c_{3}\kappa^{2}2^{-\beta l}
\).    The co\textsf{}nditions of  Proposition~\ref{pr:unbiasedML} are thus met for \(Y=U\) and \(Y_{l}=\hat U_{l}\) for \(l\ge0\), with \(\nu=c_{3}\kappa^{2}\)  and \(c=c'\).  Thus,   \(\hat V\) is square-integrable with \(E(\hat V)=E(U)\). This implies~\eqref{eq:expectedPhiApprox}.  By~\eqref{eq:||Z||^2Prop}, 
\begin{equation*}
||\hat V||^{2}\leq \frac{20c_{3}\kappa^{2}}{1-2^{-(\beta-1)/2}},
\end{equation*}
and so \(\var(\hat V)\) is upper-bounded by a constant independent of \(m\). By~\eqref{eq:TUpperBoundProp},
  the expectation of the time  to simulate \(\hat V\) is at most  \(c'/(1-2^{-(\beta-1)/2})\). This completes the proof.
\qed
\section{Proof of Theorem~\ref{th:asianApproxBis}}
By arguments similar to those used in the proof of  Theorem~\ref{th:asianApprox},  there is a constant \(c'\) independent of \(m\) and of \(\epsilon\) such that the expected cost of computing \(\hat U_{l}-\hat U_{l-1}\) is at most \(c'2^{l}\) for \(l\geq0\). 
 Also, \(||\hat U_{0}||^{2}\leq c_{3}\kappa^{2}\) and,  for \(l\geq0\), 
\begin{equation*}
||\hat U_{l}-U||^{2}\leq c_{3}\kappa^{2}2^{- l}.
\end{equation*}    The co\textsf{}nditions of  Proposition~\ref{pr:biasedML} are thus met for \(Y=U\) and \(Y_{l}=\hat U_{l}\) for \(l\ge0\), with \(\nu=c_{3}\kappa^{2}\)  and \(c=c'\).   By~\eqref{eq:SquaredBiasProp},   \(\hat V\) is square-integrable and  \((E(\hat V-U))^{2}\leq c_{3}\kappa^{2}\epsilon^{2}\). This implies~\eqref{eq:expectedPhiBias}.    Similarly,~\eqref{eq:||Z||^2PropBiased} implies that
\begin{equation*}
\var(\hat V)\leq48c_{3}\kappa^{2}\log_{2}(1/\epsilon).
\end{equation*}
Furthermore, the expectation of the time required to simulate \(\hat V\) is at most  \(4c'\log_{2}(1/\epsilon)\). \qed 
}
{
\section*{Acknowledgments}
This research has been presented at the  35th Spring International Conference of the French Finance Association, May 2018.
The author thanks Aur\'elien Alfonsi,  Mike Giles, Benjamin Jourdain  and conference participants   for helpful conversations.
This work was achieved through the Laboratory of Excellence on
Financial Regulation (Labex ReFi) under the reference ANR-10-LABX-0095. It benefitted from
a French government support managed by the National Research Agency (ANR).
\commentt{}
{
\end{APPENDIX}
\bibliography{poly}
}
\begin{table}[h]
\caption{Pricing average price calls in the Black-Scholes model with strike \(K=2\)}
\begin{footnotesize}
\begin{tabular}{llrrcrrr}\hline
    &     & $n$& Price & Std & Cost& Cost $\times$ Std$^2$ &   VRF \\ \hline
$m=125$ &RMLMC & $ 1\times10^9$ & $0.35239$ & $4.6\times10^{-5}$ & $2.1\times10^9$ & $4.5$ & $12$\\
 &MLMC & $ 8\times10^5$ & $0.35231$ & $4.6\times10^{-5}$ & $2.16\times10^9$ & $4.6$ & $12$\\  &RMLMC-Milstein & $ 1\times10^9$ & $0.35236$ & $4.4\times10^{-5}$ & $3.3\times10^9$ & $6.4$ & $8.5$\\
$m=250$ &RMLMC & $ 1\times10^9$ & $0.35126$ & $4.7\times10^{-5}$ & $2.13\times10^9$ & $4.7$ & $24$\\
 &MLMC & $ 4\times10^5$ & $0.35128$ & $4.7\times10^{-5}$ & $2.21\times10^9$ & $4.8$ & $23$\\  &RMLMC-Milstein & $ 1\times10^9$ & $0.35127$ & $4.5\times10^{-5}$ & $3.33\times10^9$ & $6.6$ & $17$\\
$m=500$ &RMLMC & $ 1\times10^9$ & $0.3507$ & $4.7\times10^{-5}$ & $2.15\times10^9$ & $4.8$ & $45$\\
 &MLMC & $ 2\times10^5$ & $0.35069$ & $4.7\times10^{-5}$ & $2.22\times10^9$ & $5$ & $43$\\
 &RMLMC-Milstein & $ 1\times10^9$ & $0.35082$ & $4.5\times10^{-5}$ & $3.36\times10^9$ & $6.8$ & $32$\\
 \hline\end{tabular}
 \end{footnotesize}
 \label{tab:BS}
\end{table}
\begin{table}
\caption{Pricing average strike calls in the Black-Scholes model}
\begin{footnotesize}
\begin{tabular}{llrrcrrr}\hline
    &     & $n$& Price & Std & Cost& Cost $\times$ Std$^2$ &   VRF \\ \hline
$m=125$ &RMLMC & $ 1\times10^9$ & $0.36325$ & $6.2\times10^{-5}$ & $1.42\times10^9$ & $5.4$ & $17$\\
 &MLMC & $ 8\times10^5$ & $0.36327$ & $4.3\times10^{-5}$ & $2.11\times10^9$ & $3.9$ & $23$\\
 &RMLMC-Milstein & $ 1\times10^9$ & $0.36332$ & $6.2\times10^{-5}$ & $2.62\times10^9$ & $10$ & $8.9$\\
$m=250$ &RMLMC & $ 1\times10^9$ & $0.36284$ & $6.3\times10^{-5}$ & $1.42\times10^9$ & $5.6$ & $34$\\
 &MLMC & $ 4\times10^5$ & $0.36291$ & $4.4\times10^{-5}$ & $2.17\times10^9$ & $4.1$ & $46$\\  &RMLMC-Milstein & $ 1\times10^9$ & $0.36291$ & $6.3\times10^{-5}$ & $2.62\times10^9$ & $11$ & $18$\\
$m=500$ &RMLMC & $ 1\times10^9$ & $0.3627$ & $6.3\times10^{-5}$ & $1.42\times10^9$ & $5.7$ & $61$\\
 &MLMC & $ 2\times10^5$ & $0.36275$ & $4.4\times10^{-5}$ & $2.13\times10^9$ & $4.2$ & $83$\\  &RMLMC-Milstein & $ 1\times10^9$ & $0.36276$ & $6.4\times10^{-5}$ & $2.63\times10^9$ & $11$ & $32$\\
 \hline\end{tabular}
 \end{footnotesize}
 \label{tab:BSAvgStrike}
\end{table}
\begin{table}
 \caption{Randomized multilevel Monte Carlo pricing of Asian calls  in the Black-Scholes model}
\begin{footnotesize}
\begin{tabular}{llrrcrr}\hline
    & $m$    & $n$& Price & Std & Cost& Cost $\times$ Std$^2$ \\ \hline
Average price &$10^7$ & $ 10^9$ & $0.35014$ & $4.8\times10^{-5}$ & $2.21\times10^9$ & $5.1$ \\
Average strike &$10^{7}$ & $ 10^9$ & $0.36252$ & $6.5\times10^{-5}$ & $1.43\times10^9$ & $6$ \\
 \hline
 \end{tabular}
\\
 The strike of the average price call is \(K=2\).
 \end{footnotesize}
 \label{tab:BSAvgPriceStrike}
 \end{table}
\begin{table}
\caption{Pricing average price  calls in Merton's jump-diffusion model with \(K=2\)}
\begin{footnotesize}
\begin{tabular}{llrrcrrr}\hline
    &     & $n$& Price & Std & Cost& Cost $\times$ Std$^2$ &   VRF \\ \hline
$m=125$ &RMLMC & $ 1\times10^9$ & $0.19306$ & $1.6\times10^{-5}$ & $2.1\times10^9$ & $0.53$ & $13$\\
 &MLMC & $ 8\times10^5$ & $0.19309$ & $1.6\times10^{-5}$ & $2.16\times10^9$ & $0.53$ & $13$\\
$m=250$ &RMLMC & $ 1\times10^9$ & $0.1924$ & $1.6\times10^{-5}$ & $2.13\times10^9$ & $0.55$ & $26$\\
 &MLMC & $ 4\times10^5$ & $0.19242$ & $1.6\times10^{-5}$ & $2.2\times10^9$ & $0.56$ & $25$\\
$m=500$ &RMLMC & $ 1\times10^9$ & $0.19206$ & $1.6\times10^{-5}$ & $2.15\times10^9$ & $0.56$ & $50$\\
 &MLMC & $ 2\times10^5$ & $0.19208$ & $1.6\times10^{-5}$ & $2.21\times10^9$ & $0.57$ & $49$\\
\hline\end{tabular}
 \end{footnotesize}
 \label{tab:MJD}
\end{table}
\begin{table}
\caption{Pricing average strike calls in Merton's jump-diffusion model}
\begin{footnotesize}
\begin{tabular}{llrrcrrr}\hline
    &     & $n$& Price & Std & Cost& Cost $\times$ Std$^2$ &   VRF \\ \hline
$m=125$ &RMLMC & $ 1\times10^9$ & $0.20107$ & $2.2\times10^{-5}$ & $1.42\times10^9$ & $0.69$ & $13$\\
 &MLMC & $ 8\times10^5$ & $0.20109$ & $1.5\times10^{-5}$ & $2.14\times10^9$ & $0.49$ & $19$\\
$m=250$ &RMLMC & $ 1\times10^9$ & $0.20096$ & $2.2\times10^{-5}$ & $1.42\times10^9$ & $0.71$ & $25$\\
 &MLMC & $ 4\times10^5$ & $0.20097$ & $1.5\times10^{-5}$ & $2.18\times10^9$ & $0.51$ & $35$\\
$m=500$ &RMLMC & $ 1\times10^9$ & $0.20088$ & $2.3\times10^{-5}$ & $1.42\times10^9$ & $0.72$ & $49$\\
 &MLMC & $ 2\times10^5$ & $0.20087$ & $1.6\times10^{-5}$ & $2.13\times10^9$ & $0.51$ & $69$\\
\hline\end{tabular}
 \end{footnotesize}
 \label{tab:MJDAvgStrike}
\end{table}
\begin{table}
\caption{Pricing average price  calls with strike  \(K=2\)  in the Square-Root diffusion model}
\begin{footnotesize}
\begin{tabular}{llrrcrrr}\hline
    &     & $n$& Price & Std & Cost& Cost $\times$ Std$^2$ &   VRF \\ \hline
$m=125$ &RMLMC & $ 1\times10^9$ & $0.21837$ & $2.0\times10^{-5}$ & $2.1\times10^9$ & $0.82$ & $13$\\
 &MLMC & $ 8\times10^5$ & $0.21839$ & $2.0\times10^{-5}$ & $2.16\times10^9$ & $0.83$ & $13$\\
$m=250$ &RMLMC & $ 1\times10^9$ & $0.21762$ & $2.0\times10^{-5}$ & $2.13\times10^9$ & $0.85$ & $26$\\
 &MLMC & $ 4\times10^5$ & $0.21763$ & $2.0\times10^{-5}$ & $2.21\times10^9$ & $0.87$ & $25$\\
$m=500$ &RMLMC & $ 1\times10^9$ & $0.21726$ & $2.0\times10^{-5}$ & $2.15\times10^9$ & $0.87$ & $50$\\
 &MLMC & $ 2\times10^5$ & $0.21728$ & $2.0\times10^{-5}$ & $2.22\times10^9$ & $0.9$ & $49$\\
\hline\end{tabular}
 \end{footnotesize}
 \label{tab:SQR}
\end{table}

\begin{table}
\caption{Pricing average strike  calls in the Square-Root diffusion model}
\begin{footnotesize}
\begin{tabular}{llrrcrrr}\hline
    &     & $n$& Price & Std & Cost& Cost $\times$ Std$^2$ &   VRF \\ \hline
$m=125$ &RMLMC & $ 1\times10^9$ & $0.2251$ & $2.9\times10^{-5}$ & $1.42\times10^9$ & $1.2$ & $11$\\
 &MLMC & $ 8\times10^5$ & $0.22505$ & $2.0\times10^{-5}$ & $2.15\times10^9$ & $0.82$ & $16$\\
$m=250$ &RMLMC & $ 1\times10^9$ & $0.22495$ & $2.9\times10^{-5}$ & $1.42\times10^9$ & $1.2$ & $21$\\
 &MLMC & $ 4\times10^5$ & $0.22485$ & $2.0\times10^{-5}$ & $2.2\times10^9$ & $0.86$ & $30$\\
$m=500$ &RMLMC & $ 1\times10^9$ & $0.22484$ & $3.0\times10^{-5}$ & $1.42\times10^9$ & $1.3$ & $40$\\
 &MLMC & $ 2\times10^5$ & $0.22483$ & $2.0\times10^{-5}$ & $2.15\times10^9$ & $0.87$ & $59$\\
\hline\end{tabular}
 \end{footnotesize}
 \label{tab:SQRAvgStrike}
\end{table}
}

\commentt{}
{
\ECSwitch
\ECHead{Supplementary Material}
}
\section{The Euler and Milstein schemes}
\label{se:EulerMil}
We show here that A2 holds when the forward price follows a  continuous diffusion
  process satisfying certain regularity conditions. Assume that  \(F(t)\)  satisfies the SDE \begin{equation*}
d F(t)=b(F(t),t)dW,
\end{equation*}   where \(b\) is a real-valued function on \(\mathbb{R}^{2}\) and  \(W\) is a one-dimensional Brownian motion under \(Q\). For \( J\subseteq\{1,\dots,m\}\), let  \(0=\tau_{0}<\tau_{1}<\cdots<\tau_{n}\) be the elements of the time grid \begin{displaymath}
G(J,l)=\{t_{j}:j\in J\} \cup \{i2^{-l}T:0\leq i\leq2^{l}\}.
\end{displaymath}Note that \( n\le|J|+2^{l}\) and the maximum distance \(\delta\) between two consecutive elements of \(G(J,l)\) is  at most \(2^{-l}T\). Using the time grid \(G(J,l)\), the Euler scheme approximates the forward price path   via the sequence \(\tilde F=\tilde F(J,l)\) defined recursively as follows:   \(\tilde F_{0}=F_{0}\) and, for \(0\leq k\leq n-1\),
\begin{equation}\label{eq:EulerRec}
\tilde F_{k+1}=\tilde F_{k}+b(\tilde F_{k},\tau_{k})(\Delta W),
\end{equation}   where \(\Delta W=W(\tau_{k+1})-W(\tau_{k})\). It follows from~\cite[Theorem 10.6.3]{kloedenPlaten1992} that, under certain regularity conditions on \(b\), 
\begin{equation}\label{eq:eulerApprox}
E(\max_{0\leq k\leq n}(\tilde F_{k}-F(\tau_{k}))^{2})\leq K_{1}\delta,
\end{equation} where \(K_{1}\) is a constant that does not depend on \(\delta\). Define \(\hat F=\hat F(J,l)\in\mathbb{R}^{J}\) as follows. For \(j\in J\), set \(\hat F_{j}=\tilde F_{k}\). where \(k\) is the index such that \(\tau_{k}=t_{j}\). In other words, \(\hat F\) is the ``restriction'' of \(\tilde F\) to the dates corresponding to \(J\). It follows from~\eqref{eq:eulerApprox}  that \(||\hat F_{j}-F_{j}||^{2}\leq K_{1}2^{- l}T\) for \(j\in J\).  Furthermore, for \(l\geq1\) and \(J'\subseteq J\subseteq\{1,\dots,m\}\), the grid \(G(J',l-1)\)  is contained in \(G(J,l)\). The vector   \((\hat F(J',l-1),\hat F(J,l))\)  can thus be simulated in at most  \(c_{1}(|J|+2^{l})\) time, where \(c_{1}\) is a constant independent of \(m\), by first simulating \(W\) on the elements of \(G(J,l)\) and then using the same \(W\) to calculate recursively \(\tilde F(J,l)\)  and \(\tilde F(J',l-1)\) via~\eqref{eq:EulerRec}. Thus A2 holds for these processes with
\(\beta=1\) for the Euler scheme. 

Similarly, under regularity conditions on \(b\), we can  calculate \(\hat F(J,l)\)  by computing the sequence \(F^*= F^*(J,l)\)  via the Milstein scheme\begin{equation*}
 F^*_{k+1}=F^*_{k}+b(F^*_{k},\tau_{k})(\Delta W)+\frac{1}{2}b(F^*_{k},\tau_{k})b'(F^*_{k},\tau_{k})((\Delta W)^{2}-(\tau_{k+1}-\tau_{k})),
\end{equation*} where \(b'\) is the partial derivative of \(b\) with respect to its first argument.  It  follows from~\cite[Theorem 10.6.3]{kloedenPlaten1992} that, under certain regularity conditions on \(b\), 
\begin{equation*}
E(\max_{0\leq k\leq n}(F^*_{k}-F(\tau_{k}))^{2})\leq K_{2}\delta^{2},
\end{equation*}where \(K_{2}\) does not depend on \(\delta\). By arguments similar to those used in the Euler scheme analysis, we conclude  that A2 holds for the Milstein scheme with \(\beta=2\)  for scalar continuous processes satisfying certain regularity conditions. A straightforward generalization of the preceding arguments shows that A2 holds for the Euler scheme with \(\beta=1\)  for multi-dimensional continuous processes satisfying certain regularity conditions.
\section{Simulation of Square-Root diffusions}
\label{se:simulationOfBESQ0}
Proposition~\ref{pr:SQRsimul} below shows how to sample \(F(t)\), for \(t\in[0,T]\).  Proposition~\ref{pr:SQRsimul} and its proof are  inspired from the analysis of the  Cox-Ingersoll-Ross  process in~\cite[\S3.4.1]{glasserman2004Monte}. \begin{proposition}\label{pr:SQRsimul}
Let \(N\) be a Poisson random variable with mean \(2F_{0}/(\sigma^{2}t)\). For integer \(k\geq1\), let   \(\chi^{2}_{k}\)  be a Chi-Square  random variable with \(k\) degrees of freedom independent of \(N\), and let    \(\chi^{2}_{0}=0\).  Then \(F(t)\) has the same distribution as \((\sigma^{2}t/4)\chi^{2}_{2N}\).
Furthermore, \(F(t)\) is square-integrable.\end{proposition}
\commentt{\begin{proof}}{\proof{Proof.}}
For \(t\in[0,T]\), let \(X(t)=4F(t)/\sigma^{2}\),  and let \(x=X(0)\). Then \begin{equation}\label{eq:BesselSDE}
X(t)=x+2\int^{t}_{0} \sqrt{X(s)}\, dW(s).
\end{equation}
Hence \(X\) is a squared Bessel process of dimension \(0\). Such a process is a martingale~\cite[p.~339]{jeanblanc2009mathematical}, and so \(\int^{t}_{0} X(s)\, ds\) has  finite expectation. By~\eqref{eq:BesselSDE} and the isometry of stochastic integrals~\cite[\S1.5.1]{jeanblanc2009mathematical}, it follows that  \(X(t)\) is square-integrable.   By~\cite[p.~344]{jeanblanc2009mathematical}, for \(t>0\),   we have  \(\Pr(X(t)=0)=e^{-x/(2t)}\)  and \(X(t)\) has density\begin{displaymath}
q_{t}(x,y)=\frac{1}{2t}\sqrt{\frac{x}{y}}\exp(-\frac{x+y}{2t})I_{1}(\frac{\sqrt{xy}}{t})
\end{displaymath} at \(y>0\), where \(I_{1}\) is the modified Bessel function with index \(1\) defined for \(z>0\) by\begin{displaymath}
I_{1}(z)=\sum^{\infty}_{k=0}\frac{(z/2)^{2k+1}}{k!(k+1)!}.
\end{displaymath}
For \(y>0\) and \(k\geq1\),  \begin{displaymath}
\Pr(\chi^{2}_{2k}\geq y)=\frac{1}{2}\int^{\infty}_{y} e^{-z/2}\frac{(z/2)^{k-1}}{(k-1)!}\,dz.
\end{displaymath} 
Thus,  \begin{displaymath}
\Pr(t\chi^{2}_{2k}\geq y)=\frac{1}{2t}\int^{\infty}_{y} \exp(-\frac{z}{2t})(\frac{z}{2t})^{k-1}\frac{1}{(k-1)!}\,dz.
\end{displaymath}Since \(E(N)=x/(2t)\), we have \begin{displaymath}
\Pr(N=k)=\exp(-\frac{x}{2t})(\frac{x}{2t})^{k}\frac{1}{k!},
\end{displaymath} and so \begin{eqnarray*}
\Pr(t\chi^{2}_{2N}\geq y)&=&\sum_{k=1}^\infty \Pr(N=k)\Pr(t\chi^{2}_{2k}\geq y)
\\&=&\frac{1}{2t}\int^{\infty}_{y}\exp(-\frac{x+z}{2t})\sum^{\infty}_{k=1}(\frac{x}{2t})^{k}(\frac{z}{2t})^{k-1}\frac{1}{(k-1)!k!}\,dz\\
&=&\int^{\infty}_{y} q_{t}(x,z)\,dz\\&=&\Pr(X(t)\geq y).\end{eqnarray*}
Thus,  \(X(t)\) has the same distribution as \(t\chi^{2}_{2N}\).
This concludes the proof. 
\commentt{\end{proof}}{\Halmos\endproof} 
Consider now a time grid \(G\)  consisting of \(n+1\) dates \(0=\tau_{0}<\tau_{1}<\cdots<\tau_{n}\). We can use Proposition~\ref{pr:SQRsimul} to recursively sample \(F(\tau_{k})\) for \(1\leq k\leq n\), and thereby simulate the forward price process   on \(G\) in \(O(n)\) expected time.
Algorithms that simulate in unit expected time Poisson and Chi-Square random variables are given in~\cite{DevroyeSpringer}. In our experiments, though, we have used generators  from the standard   C++ library. 
\section{Further numerical experiments}
We report here additional numerical experiments for the Black-Scholes model, Merton's jump-diffusion model, and the Square-Root diffusion model, using the same model parameters as in~\S\ref{se:numericalExamples}.   
\subsection{The Black-Scholes model}Table~\ref{tab:BSVRFs} gives VRFs  for average price calls  with different strikes for the  RMLMC, MLMC and RMLMC-Milstein algorithms, with  \(S_0=2\), \(\sigma=50\%\), 
 \(r=5\%\), and \(T=2\).   For each strike, the VRFs are proportional to \(m\) for the three algorithms. The RMLMC and MLMC methods have a similar performance, and  slightly outperform the  RMLMC-Milstein algorithm.
 
\begin{table}
\caption{Variance reduction factors for average price calls in the Black-Scholes model}
\begin{footnotesize}
\begin{tabular}{llccccc}\hline
$K$    &     & $1.6$& $1.8$ & $2$ & $2.2$& $2.4$ \\ \hline
$m=125$ &  RMLMC & $ 13$ & $13$ & $12$ & $11$ & $11$ \\
&MLMC&$13$&$12$&$12$&$11$&$10$\\
 &  RMLMC-Milstein & $ 8.9$ & $8.7 $ & $8.4$ & $8.2$ & $8$ \\
$m=250$ &  RMLMC & $ 26$ & $25$ & $23$ & $22$ & $21$ \\
&MLMC&$25$&$ 24$ & $22$ & $22$ & $22$\\
 &  RMLMC-Milstein & $ 17$ & $17$ & $17$ & $16$ & $16$ \\
$m=500$ &  RMLMC & $ 49$ & $47$ & $45$ & $43$ & $41$ \\
&MLMC&$ 49$&$44$& $44$& $43$& $39$\\ 
&  RMLMC-Milstein & $ 34$ & $33$ & $32$ & $31$ & $30$\\
\hline\end{tabular}
\\  \(n=10^{7}\) for the RMLMC and RMLMC-Milstein algorithms and \(n=10^{6}/m\) for the MLMC algorithm.\end{footnotesize}
\label{tab:BSVRFs}
\end{table}
\subsection{Merton's jump-diffusion model}
 Table~\ref{tab:MJDAvgPriceStrike} gives prices of average price and average strike calls
when \(m=10^{7}\) using the RMLMC algorithm, with   \(S_0=2\),  \(\sigma=17.65\%\),   $r = 5.59\%$, $q = 1.14\%$, \(\lambda= 8.90\%\), \(\beta=-88.98\%\), \(\gamma=45.05\%\), and \(T=2\).
\begin{table}
\caption{Randomized Multilevel Monte Carlo pricing of Asian calls  in Merton's jump-diffusion model}
\begin{footnotesize}
\begin{tabular}{rrccr}\hline
    &      Price & Std & Cost& Cost $\times$ Std$^2$ \\ \hline
Average price &  $0.19173$ & $1.6\times10^{-5}$ & $2.21\times10^9$ & $0.6$ \\
Average strike  & $0.20082$ & $2.3\times10^{-5}$ & $1.42\times10^9$ & $0.75$\\
 \hline\end{tabular}
 \\  \(m=10^{7}\) and \(n=10^{9}\). The strike of the average price call is \(K=2\). 
 \end{footnotesize}
 \label{tab:MJDAvgPriceStrike}
\end{table}
\subsection{The Square-Root diffusion model}
Table~\ref{tab:SQRAvgPriceStrike} gives prices of average price and average strike calls
when \(m=10^{7}\) using the RMLMC algorithm, with \(S_{0}=2\), \(r=5\%\), \(\sigma=0.4\), and  \(T=2\).\begin{table}
\caption{Randomized multilevel Monte Carlo pricing of Asian calls   in the Square-Root diffusion model}
\begin{footnotesize}
\begin{tabular}{rrccr}\hline
    &      Price & Std & Cost& Cost $\times$ Std$^2$ \\ \hline
Average price &  $0.21693$ & $2.0\times10^{-5}$ & $2.21\times10^9$ & $0.92$ \\
Average strike  & $0.22474$ & $3.1\times10^{-5}$ & $1.43\times10^9$ & $1.3$\\
 \hline
 \end{tabular}
\\    \(m=10^{7}\) and \(n=10^{9}\). The strike of the average price call is \(K=2\). 
 \end{footnotesize}
 \label{tab:SQRAvgPriceStrike}
\end{table}

\commentt{
\bibliography{poly}
}{}
\end{document}